\documentclass[aps,prb,groupedaddress,nofootinbib,floatfix,twocolumn,showpacs]{revtex4}
\usepackage{graphicx}
\usepackage{amsmath,amssymb,bm,hyperref,color}
\usepackage{psfrag}
\definecolor{blue}{rgb}{0.0,0.0,1.0}
\definecolor{black}{rgb}{0.0,0.0,0.0}
\definecolor{red}{rgb}{1.0,0.0,0.0}
\hypersetup{colorlinks=true,breaklinks, linkcolor=blue,urlcolor=blue,citecolor=red}

\newcommand{\abs}[1]{\ensuremath{\lvert#1\rvert}}

\def \sgn {\mathop {\rm sgn}}
\def \Im {\mathop {\rm Im}}

\def \Tr {\mathop {\rm Tr}}

\begin{document}
\title{Superperturbation solver for quantum impurity models}

\author{Hartmut Hafermann}
\email[\hspace{-1.4mm}]{hartmut.hafermann@physnet.uni-hamburg.de}
\affiliation{I. Institute for Theoretical Physics, University of Hamburg, 20355 Hamburg, Germany }
\author{Christoph Jung}
\affiliation{I. Institute for Theoretical Physics, University of Hamburg, 20355 Hamburg, Germany }
\author{Sergey Brener}
\affiliation{I. Institute for Theoretical Physics, University of Hamburg, 20355 Hamburg, Germany }
\author{Mikhail I. Katsnelson}
\affiliation{Institute for Molecules and Materials, Radboud University of Nijmegen, 6525 AJ Nijmegen, The Netherlands }
\author{Alexei N. Rubtsov}
\affiliation{Department of Physics, Moscow State University, 119992 Moscow, Russia}
\author{Alexander I. Lichtenstein}
\affiliation{I. Institute for Theoretical Physics, University of Hamburg, 20355 Hamburg, Germany }

\date{\today}

\begin{abstract}
We present a very efficient solver for the general Anderson impurity problem. It is based on the 
perturbation around a solution obtained from exact diagonalization using a small number of bath sites. We formulate a perturbation theory which is valid for both weak and strong coupling and interpolates between these limits.
Good agreement with numerically exact quantum Monte-Carlo results is found for a single bath site over a wide range of parameters. In particular, the Kondo resonance in the intermediate coupling regime is well reproduced for a single bath site and the lowest order correction. The method is particularly suited for low temperatures and alleviates analytical continuation of imaginary time data due to the absence of statistical noise compared to quantum Monte-Carlo impurity solvers.
\end{abstract}

\maketitle

\section{Introduction}

Quantum impurity models have been widely used in condensed matter physics. Examples comprise the study of the Kondo-effect\cite{hewson}, or of adatoms on surfaces\cite{savkin}. In particular, the success of the dynamical mean-field theory (DMFT) to describe strongly correlated systems has triggered efforts to develop efficient solvers for the impurity problem. In DMFT, the lattice problem is mapped onto an effective quantum impurity problem which needs to be solved repeatedly to satisfy a self-consistency condition. 
Generalizations of the DMFT, such as cluster DMFT or dynamical cluster approximations require highly efficient methods for the solution of the multiorbital quantum impurity problem.

The combination of the DMFT with electronic structure methods, such as the local density approximation (LDA+DMFT approach)\cite{kotliar} requires the solution of impurity models with general types of the interactions, such as the full Coulomb vertex. 
The problem has been solved using approximate methods such as a multiorbital version of the fluctuation-exchange approximation (FLEX) for weak coupling\cite{katsnelson:rmp}. A strong-coupling solver based on an expansion in the impurity-bath hybridization has been proposed in order to address systems with open d or f-shells or Mott insulators\cite{dai:045111}.
Recently, next-generation Monte-Carlo methods provide a numerically exact solution\cite{rubtsovctqmc,werner1,haule}. While being suitable for general interactions, these methods however require a sizeable numerical effort and analytical continuation is complicated through statistical noise in the imaginary time data.

In this article we propose an efficient solver for the impurity problem which is suitable for both weak and strong coupling and general interactions. It is essentially based on a ``superperturbation'', i.e. a perturbation on top of a solution obtained by exact diagonalization (ED) for a small number of bath levels. The method alleviates analytical continuation due to the absence of noise and is suitable for the study of, e.g., multiplet effects in solids.

\section{Formalism}
For notational convenience, we introduce the formalism for the single-orbital case. It was introduced earlier in the context of  lattice fermion models to include spatial correlations beyond DMFT in Ref. \onlinecite{rubtsov}. A generalization of the underlying concepts to the multiorbital case can be found in Refs. \onlinecite{hafermann,brener}. In a complementary approach, named dynamical vertex approximation, the single- and two-particle Green functions of the impurity were computed using ED\cite{toschi}.
The Hamiltonian of the Anderson impurity model (AIM) reads
\begin{equation}
H_{\text{AIM}}=\sum_{p\sigma} \epsilon_p b^\dagger_{p\sigma} b_{p\sigma} + \sum_{p\sigma} V_p (b^\dagger_{p\sigma} c_\sigma + \text{h.c.}) + H_{\text{int}}[c^\dagger,c]\ ,
\end{equation}
where the local impurity degrees of freedom represented by $c^\dagger$,$c$ couple to a bath of free conduction electrons with dispersion $\epsilon_p$ via a hybridization $V_p$. $H_{\text{int}}$ stands for any local interaction. Since $H_{\text{AIM}}$ is quadratic in the bath operators $b^\dagger$,$b$, they can be integrated out giving rise to the action (henceforth we switch to the path integral representation):
\begin{equation}
S[c^\ast,c]=-\sum_{\omega\sigma} c^\ast_{\omega\sigma}\left((i\omega+\mu)-\Delta_{\omega\sigma}\right)c_{\omega\sigma}+ H_{\text{int}}[c^\ast,c]\ .
\label{eqn::impaction}
\end{equation}
Here $\mu$ is the chemical potential and the sum is over Matsubara frequencies $\omega_n=(2n+1)\pi/\beta$, where $\beta$ is the inverse temperature. In ED the continuous dispersion $\epsilon_p$ of the bath is approximated by a finite number of bath levels, which corresponds to replacing the hybridization function $\Delta$ by its discrete counterpart
\begin{equation}
\Delta^{(n)}_{\omega\sigma} = \sum_{p=1}^{n} \frac{\abs{V^\sigma_p}^2}{i\omega - \epsilon^\sigma_p}\ .
\end{equation}
The problem now is to determine the parameters $V_p$,$\epsilon_p$ such that $\Delta^{(n)}$ is the best approximation to the original hybridization $\Delta$. Mathematically, this corresponds to projecting the hybridization function onto the discrete subspace spanned by the parameters $V_p$,$\epsilon_p$. Various methods have been proposed for this purpose\cite{georges,koch}. One way is to perform a conjugate gradient minimization of, e.g., the distance function
\begin{equation}
d = \frac{1}{\omega_{\text{max}}}\sum_{\omega}^{\omega_\text{max}} \abs{\omega}^{-k}\, \abs{\, \Delta_{\omega\sigma}\, -\, \Delta^{(n)}_{\omega\sigma}\, }^2\ ,
\label{eqn::dist}
\end{equation}
where the sum is over Matsubara frequencies. The parameter $k$, if chosen large (e.g. $k=3$), enhances the importance of the lowest Matsubara frequencies in the minimization procedure. This parameter is particularly important when a small number of bath sites is used to approximate the original hybridization.

The quality of the approximation can be measured by $d$ and is determined by the number of bath sites used to represent the original hybridization. On the other hand, the number of bath sites is limited due to the exponential growth of the Hilbert space.
Instead of approximating $\Delta$ by a large number of bath sites, we shall follow a different route and rewrite the action Eqn. \ref{eqn::impaction} in terms of an exactly solvable part and an additional bilinear term:
\begin{eqnarray}
S[c^\ast,c]&=&-\sum_{\omega\sigma} c^\ast_{\omega\sigma}((i\omega+\mu)-\Delta^{(n)}_{\omega\sigma})c_{\omega\sigma} 
+ H_{\text{int}}[c^\ast,c]\nonumber\\
&& \qquad - \sum_{\omega\sigma} c^\ast_{\omega\sigma} (\Delta^{(n)}_{\omega\sigma} -\Delta_{\omega\sigma}) c_{\omega\sigma}\ .
\label{eqn::action_rew}
\end{eqnarray}
The exactly solvable part (the first line of Eqn. \ref{eqn::action_rew}) we will henceforth refer to as $S^{(n)}$.
Clearly, the difference $D_{\omega\sigma}=\Delta^{(n)}_{\omega\sigma}-\Delta_{\omega\sigma}$ can be made arbitrarily small by including more and more bath sites. The main point is that the number of bath sites which we will employ to solve (\ref{eqn::action_rew}) and hence the Hilbert space is much smaller than for the case of conventional ED. Since $S^{(n)}$ in general is non-Gaussian, Wick's theorem is not directly applicable. We formulate a perturbation theory in $D$ by introducing new fermionic degrees of freedom in the path integral for the partition function by means of the following identity:
\begin{eqnarray}
&&\!\!\!\!\int\exp \left(-f^*[g_{\omega\sigma}D g_{\omega\sigma}]^{-1} f - f^*g_{\omega\sigma}^{-1}c - c^*g_{\omega\sigma}^{-1}f \right) \mathcal{D}[f^*,f]\nonumber\\
&&\quad =\det(g_{\omega\sigma}D_{\omega\sigma}g_{\omega\sigma})^{-1}\exp\left(c^* D_{\omega\sigma} c\right)\ .
\label{eqn::gaussid}
\end{eqnarray}
A similar approach has been used to derive a strong coupling expansion for the Hubbard model\cite{tremblay}. In the above equation $g_{\omega\sigma}\equiv g_{\omega\sigma}^{(n)}\equiv -\langle c_{\omega\sigma}c^*_{\omega\sigma}\rangle^{(n)}$ denotes the impurity Green function w.r.t. the discrete hybridization $\Delta^{(n)}$, which can be calculated exactly.
With this substitution, the action becomes
\begin{eqnarray}
S[c^*,c;f^*,f]=S^{(n)} &+& \sum_{\omega\sigma} f^*_{\omega\sigma} g_{\omega\sigma}^{-1} c_{\omega\sigma} + c^*_{\omega\sigma} g_{\omega\sigma}^{-1} f_{\omega\sigma}\nonumber\\
&& + \sum_{\omega\sigma} f^*_{\omega\sigma} g_{\omega\sigma}^{-1}D_{\omega\sigma}g_{\omega\sigma}^{-1} f_{\omega\sigma}\ .
\end{eqnarray}

\begin{figure}[t!]
\begin{center}
\psfrag{s}{$G^f$}
\psfrag{g}{$\Sigma^f$}
\psfrag{x}[c]{$\ \ \ G^{f\,-1}=G_0^{f\,-1}-\Sigma^f$}
\includegraphics[scale=0.25,angle=0]{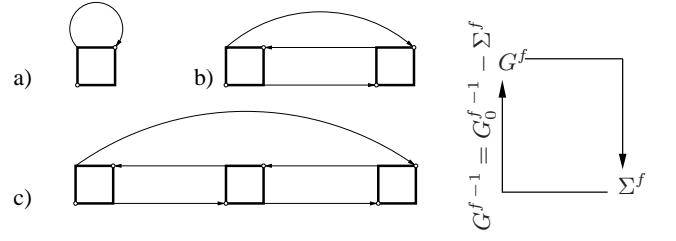}
\end{center}
\caption{Lowest order diagrams for the self-energy $\Sigma^f$ with the action given by Eqn. \ref{eqn::sf} and illustration of the Dyson iterations: The self-energy is obtained from $G^f$ by summing up the diagrams a)-c). From Dyson's equation, we in turn obtain $G^f$, which is subsequently used in the diagrams.}
\label{fig::diagrams}
\end{figure}

From the expression for the partition function $\mathcal{Z}=\mathcal{Z}^{(n)}\int\int$ $\exp(-S[c^*,c;f^*,f])\mathcal{D}[c^*,c]\mathcal{D}[f^*,f]$, we now integrate out $c^*$ and $c$ by expanding the  exponential $\exp(-\sum_{\omega\sigma} f^*_{\omega\sigma} g_{\omega\sigma}^{-1} c_{\omega\sigma} + c^*_{\omega\sigma} g_{\omega\sigma}^{-1} f_{\omega\sigma})$ and using the fact that the functional integral over $\exp(-S^{(n)})$ produces correlation functions that can be obtained exactly from the ED, such as the two-particle Green function:
\begin{equation}
\chi^{(n)}_{1234} = \frac{1}{\mathcal{Z}^{(n)}}\int c_1 c_2^* c_3 c_4^* \exp(-S^{(n)})\mathcal{D}[c^*,c]\ ,
\end{equation}
where we have used the shorthand notation $1\equiv\{\omega_1\sigma_1\}$. A compact expression for the Lehmann representation of the two-particle Green function is given in appendix \ref{app::a}.
Here we carry out this expansion up to fourth order (note that odd terms drop out of the expansion), with the result
\begin{equation}
S^f[f^*,f]=-\sum_{\omega\sigma}f^*_{\omega\sigma}(G_0^{f})^{-1}_{\omega\sigma} f_{\omega\sigma} + \gamma^{(n)}_{1234}\, f^*_1 f_2 f^*_3 f_4\ .
\label{eqn::sf}
\end{equation}
Here $\gamma^{(n)}$ is the two-particle vertex constructed from the two-particle Green function
as (henceforth we omit the superscript ``$(n)$'' on $\chi$ and $\gamma$)
\begin{equation}
\gamma_{1234}=g_{11'}^{-1}g_{33'}^{-1}(\chi_{1'2'3'4'}-\chi_{1'2'3'4'}^0)g_{2'2}^{-1}g_{3'3}^{-1}\ ,
\label{eqn::gammanb}
\end{equation} 
with $\chi^0_{1234}=\beta (g_{12}g_{34}-g_{14}g_{32})$. 

The auxiliary Green function $G^f$ is given by
\begin{equation}
G^{f}_{0\,\omega\sigma} = -g_{\omega\sigma}(g_{\omega\sigma}+(\Delta^{(n)}_{\omega\sigma}-\Delta_{\omega\sigma})^{-1})^{-1}g_{\omega\sigma}\ .
\label{eqn::gf}
\end{equation}
This function has some remarkable properties. Let us discuss these for the case $\Delta^{(n)}=0$ and Hubbard interaction $H_{\text{int}}=U n_\uparrow n_\downarrow$. Then we have $D=-\Delta$ and we can approximate $G^f$ for strong hybridization by $G\approx g$. In the weak coupling limit, i.e. $U\rightarrow 0$, $g$ approaches the bare Green function and $\gamma\sim U$, so that the dual perturbation theory becomes equivalent to conventional perturbation theory. On the other hand, for $\Delta$ small, $G^f$ itself is small and can obviously be approximated as $G^f\approx g_{\omega\sigma}\Delta_{\omega\sigma}g_{\omega\sigma}$.  This generates a fast converging strong coupling perturbation expansion around the atomic limit. In fact, one can show that in this case the expression for Green's function obtained by a hybridization expansion of the Green function given in Ref. \onlinecite{dai:045111} is contained in the lowest order diagram of our expansion. To see this, we write the auxiliary Green function as $G^f\approx G^f_0 + G^f_0 \Sigma^{f}_{(a)} G^f_0$, where the self-energy correction of diagram a) in Fig. \ref{fig::diagrams} is given by $\Sigma^{f}_{(a)}=-\gamma_{1234}(G^f_0)_{43}$. In order to compare this to known results, we have to relate the auxiliary Green function to the Green function of the impurity, $G_{12}=-\langle c_1 c_2^*\rangle$. The fact that we have used an exact identity to introduce the auxiliary fermions allows us to establish an exact relation between $G$ and $G^f$, i.e.
\begin{equation}
 G_{\omega\sigma} = D_{\omega\sigma}^{-1} + (g_{\omega\sigma}D_{\omega\sigma})^{-1} G_{\omega\sigma}^f (D_{\omega\sigma}G_{\omega\sigma})^{-1}
\end{equation}
(see Ref. \onlinecite{rubtsov}).
\begin{figure}[t]
\psfrag{x}{$\omega$}
\psfrag{y}[c]{$\Im G(i\omega)$}
\psfrag{z}[c]{$\Im \Delta(i\omega)$}
\begin{center}
\includegraphics[scale=0.65,angle=0]{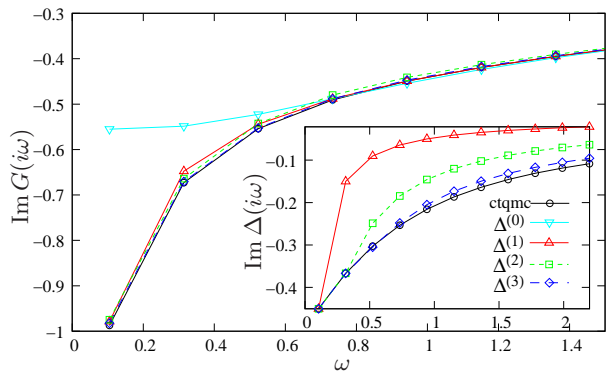}
\end{center}
\caption{(Color online) Imaginary part of the impurity Green function obtained by the superperturbation using different numbers of bath sites for $\beta=30$ and $U=3$ The representation of the exact hybridization by $\Delta^{(n)}$ is shown in the inset for $n>0$ ($\Delta^{(0)}\equiv 0$).}
\label{fig::comp_delta}
\end{figure}
\begin{figure}[t]
\psfrag{x}{$\omega$}
\psfrag{y}[c]{$\Im G(i\omega)$}
\begin{center}
\includegraphics[scale=0.65,angle=0]{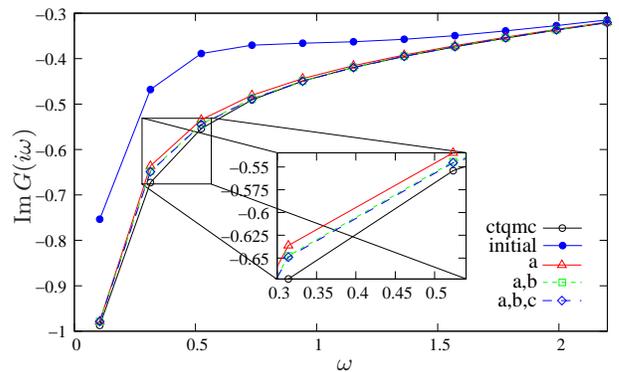}
\end{center}
\caption{(Color online) The contribution of different diagrams to the superperturbation result compared to the exact solution (open circles), for a single bath site. The paramters are otherwise the same as in Fig. \ref{fig::comp_delta}. Diagram a) yields by far the largest correction (upward triangles) to the initial solution obtained from ED (filled circles).}
\label{fig::comp_diags}
\end{figure}
Inserting the approximation to $G^f$ into this equation yields the following expression for $G$ after some straightforward algebra:
\begin{eqnarray}
G_{12} &=& [g(Dg + 1)^{-1}]_{12} - [(1+g D)^{-1}]_{11'}\times\nonumber\\
&\times& (\chi-\chi^0)_{1'2'3'4'} [(g+D^{-1})^{-1}]_{4'3'} [(Dg+1)^{-1}]_{2'2},\nonumber\\
\label{eqn::G}
\end{eqnarray}
where we have used that fact that the vertex $\gamma$ is expressed in terms of the two-particle Green function by Eqn. \ref{eqn::gammanb}. Considering the first term in the expression for $\chi^0$ yields a contribution $-\beta g_{12}g_{34}[(g+D^{-1})^{-1}]_{43} = -\beta \Tr [g(g+D^{-1})^{-1}]$. In order to compare with Ref. \onlinecite{dai:045111}, we take $D=-\Delta$, as above. In this case, $g$ corresponds to the Green function for the atomic limit. For small $\Delta$ we have $(1-\Delta g)^{-1}\rightarrow 1$ and $(g-\Delta^{-1})^{-1}\rightarrow -\Delta$. Gathering the results we can approximate $G$ 
in the limit of small $\Delta$ as
\begin{equation}
G_{12} \approx g_{12} + g_{12}\, \beta\, \Tr[g\,\Delta] + \chi_{1234}\Delta_{43} \ ,
\end{equation}
which is essentially Eqn. (13) of Ref. \onlinecite{dai:045111}. 

Hence the dual perturbation theory has the correct limiting behavior in both and weak and strong coupling limits. It therefore interpolates between these limits, so that sensible results can be anticipated even in the intermediate coupling regime. In the following we will demonstrate that this indeed is the case. For intermediate coupling, we further exploit the possibility to improve the starting point of the perturbation theory by expanding around the ED solution for a finite number of bath sites. In this case, low energy Kondo physics and the high energy incoherent features are correctly reproduced.

\section{Results}

The results shown in the following were performed using the diagrams a) to c) depicted in Fig. \ref{fig::diagrams}. If not otherwise stated, the results were obtained by making use of the Dyson iterations illustrated in the same figure: The self-energy was calculated using the bare auxiliary Green function, Eqn. \ref{eqn::gf}, on the first iteration. Inserting the self-energy into the Dyson equation yields a new Green function which is subsequently used in the diagrams on the next iteration. This procedure is carried out until self-consistency and converges in typically less than ten iterations.

In the following we will discuss results obtained for the case of Hubbard interaction $H_{\text{int}}=U \int_0^\beta d\tau n_\uparrow(\tau)n_\downarrow(\tau)$. In order to test our approach, we performed calculations for up to $n=3$ bath sites. The calculations were done for a semielliptical density of states of bandwidth $W=4t$, with $\Delta_\omega=-2i t^2/(\omega+\sqrt{\omega^2+4t^2})$. We take the half-bandwidth as the energy unit: $W/2=1$.

In Fig. \ref{fig::comp_delta} we present results for $U=3$, obtained for different number of bath sites up to $n=3$. The quality of the representation of $\Delta$ by $\Delta^{(n)}$ is shown in the inset for $n>0$. In order to determine the parameters $V_p$ and $\epsilon_p$, we have minimized the distance function, Eqn. \ref{eqn::dist} for $k=3$. This choice enhances the importance of the lowest frequencies in the minimization procedure. As can be seen in the inset, this condition results in $\Delta$ and $\Delta^{(n)}$ being equal on the first $n$ Matsubara frequencies. This turns out to be a good starting point for the perturbation theory. Using a smaller $k$ leads to a $\Delta^{(n)}$ which better represents the tail of the hybridization function and generally leads to worse results.

The superperturbation results are compared to those of a numerically exact continuous-time quantum Monte-Carlo (CTQMC) calculation. One can see that while the expansion around the atomic limit lacks accuracy, a drastic improvement occurs for a single bath site, although the difference between $\Delta^{(1)}$ and $\Delta$ is still significant. For $n>1$, the results are essentially converged. We find qualitatively the same behavior for a wide-range of $U$ values. The fast convergence w.r.t the number of bath sites significantly reduces the computational effort compared to CTQMC calculations. 

In the figure we have plotted the results obtained from summing up skeleton diagrams. The use of skeleton diagrams is theoretically relevant since in this case the result is conserving in the Baym-Kadanoff sense\cite{BK}. The results obtained from the first Dyson iteration or from the lowest-order approximation, i.e. $G^f=G_0^f + G_0^f \Sigma^f G_0^f$, however, achieve the same quality of approximation (not shown here).

Let us now investigate the role of the different diagrams in the perturbation expansion. To this end, we have plotted the results for the same parameters as in the previous figure, for different combinations of the diagrams shown in Fig. \ref{fig::comp_diags}. We have used only a single bath site, for which the exact solution (open circles) and the initial solution obtained by ED (filled circles) differ substantially. One can clearly see that diagram a) yields by far the largest correction to the initial solution. The value on the first Matsubara frequency is almost exactly reproduced. This might be expected, since $\Delta^{(1)}$ is identical to $\Delta$ on the first Matsubara frequency. The largest deviations occur for the second and third frequency. We have zoomed into this region in the inset. One can see that all diagrams give a correction in the right direction, whereby the correction by diagram c) is negligible.

\begin{figure}[t]
\psfrag{x}{$\omega$}
\psfrag{y}[c]{DOS}
\begin{center}
\includegraphics[scale=0.65,angle=0]{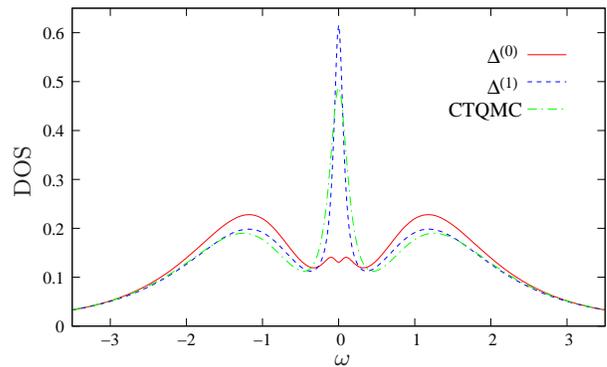}
\end{center}
\caption{(Color online) Comparison of the maximum entropy density of states obtained by superperturbation for no ($\Delta^{(0)}$) and one bath site ($\Delta^{(1)}$) to the numerically exact (continuous-time quantum Monte-Carlo) result. While the superperturbation around the atomic limit ($\Delta^{(0)}$) does not reproduce the Kondo resonance, the perturbation around the solution for one bath site already contains this physics.}
\label{fig::kondo}
\end{figure}

\begin{figure}[t]
\psfrag{x1}{$\omega$}
\psfrag{y1}[c]{DOS}
\psfrag{x2}{$\tau$}
\psfrag{y2}[c]{$G(\tau)$}
\begin{center}
\includegraphics[scale=0.65,angle=0]{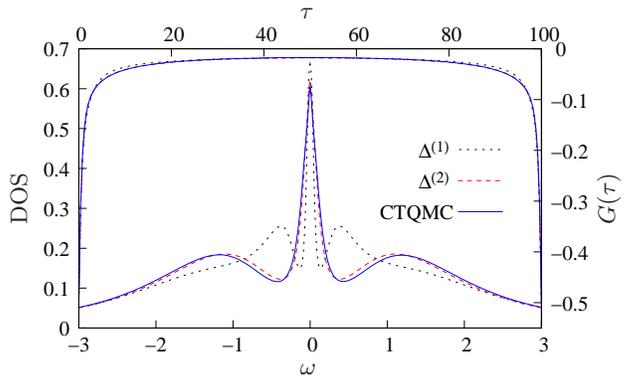}
\end{center}
\caption{(Color online) Low temperature results for $U=3$ and $\beta=100$. Shown are the imaginary time Green function (upper and right axis) and the corresponding maximum entropy density of states (lower and left axis). The result obtained by superperturbation  around two bath sites requires considerably less computational effort compared to QMC and is almost indistinguishable from the exact (CTQMC) result.}
\label{fig::lowtemp}
\end{figure}

From Fig. \ref{fig::kondo} it is obvious that also spectral properties are correctly reproduced. We show the maximum entropy density of states\cite{jarrell} (DOS) obtained from the imaginary time data. The analytical continuation of the quantum Monte-Carlo data (dashed-dotted line) exhibits the two Hubbard bands at $\omega=\pm U/2$ and shows the Kondo resonance at the Fermi level. We cannot reproduce the Kondo physics by perturbing around the atomic limit ($\Delta^{(0)}$, solid line), in accordance with the findings in Ref. \onlinecite{dai:045111}. However, perturbation around the ED solution for a single bath-site already captures the Kondo resonance and yields good agreement compared to the exact solution.

In order to demonstrate that the approach also works for low temperatures, we present results for $T=0.01$ in Fig.  \ref{fig::lowtemp}. Although the expansion around the solution for a single bath site ($\Delta^{(1)}$) shows small deviations in the imaginary time Green function $G(\tau)$, the approximation appears insufficient as seen in the density of states. The superperturbation around the two bath-site solution however is almost exact.

As already mentioned, the formalism of introducing auxiliary degrees of freedom in the path integral representation presented here has originally been introduced in the context of lattice models, termed ``dual fermion approach''\cite{rubtsov}. In that case the action is decomposed into an impurity part with hybridization $\Delta $ and a term which contains the bare dispersion of the lattice fermions. 
Since this decomposition is independent of the form of the hybridization, it can be represented by the discrete hybridization $\Delta^{(n)}$. Hence the present framework can be used to combine dual fermion calculations with ED for the efficient solution of lattice models.

\section{Conclusions}

In conclusion, we have presented an efficient approximate solver for the quantum impurity problem. The formalism straightforwardly generalizes to the multiorbital case and is suitable for a general form of the interaction. The perturbation expansion has been shown to be convergent in both weak and strong coupling limits. It was further demonstrated that the solver also works well in the intermediate coupling regime and down to low temperatures. The approximation is controlled in the sense that the final result can be judged on the basis of how well the input hybridization is approximated by a finite number of bath sites. The method can be used for the study of multiplet effects in solids in the context of realistic LDA+DMFT calculations.

\begin{acknowledgments}
We would like to thank Erik Koch for valuable discussions.
This work was supported by DFG Grant No. SFB 668-A3 (Germany), the Leading
scientific schools program and the ``Dynasty'' foundation (Russia).
\end{acknowledgments}

\appendix

\section{Lehmann representation of the two-particle Green function}
\label{app::a}
In this appendix, we derive a compact expression for the Lehmann representation of the two-particle Green function (2PGF). A similar expression was given in Ref. \cite{toschi} without explicit consideration of the singular contributions. By definition, the 2PGF in Matsubara space is given by
\begin{eqnarray}
\chi_{\omega_1 \omega_2 \omega_3}^{\sigma\sigma'}&=&\frac{1}{\beta^2}\int_0^\beta\! d\tau_1
\int_0^\beta\! d\tau_2 \int_0^\beta\! d\tau_3\, e^{i ( \omega_1\tau_1 +\omega\tau_2+\omega_3\tau_3)}\times\nonumber\\
&&\times \langle T_\tau c_{\sigma}(\tau_1)c^\dagger_{\sigma}(\tau_2)c_{\sigma'}(\tau_3)c^\dagger_{\sigma'}(0)\rangle\, \ .
\label{eqn::chi_omega}
\end{eqnarray}
Here time translation invariance of the imaginary time 2PGF has been used. Note that here the frequencies in the exponential corresponding to annihilation and creation operators have the same sign in contrast to the usual definition for the Fourier transform. Correspondingly, energy conservation requires $\omega_1+\omega_2+\omega_3+\omega_4=0$. By restricting the range of integration such that time ordering is explicit, one obtains $3!$ different terms. These can be brought into the same form by permuting the operators \emph{and} corresponding frequencies. By the anticommutation relations, each term picks up the sign of the permutation. After introducing the sum over eigenstates, the 2PGF can be written as
\begin{eqnarray}
\lefteqn{\chi_{\omega_1 \omega_2 \omega_3}^{\sigma\sigma'}\!=\!\frac{1}{\mathcal{Z}}\sum_{ijkl}\sum_{\Pi}\! \phi (E_i,E_j,E_k,E_l,\omega_{\Pi_1},\omega_{\Pi_2},\omega_{\Pi_3})}\nonumber\\
&\times&\!\!\sgn(\Pi) \langle i|\mathcal{O}_{\Pi_1}|j\rangle\,\langle j|\mathcal{O}_{\Pi_2}|k\rangle\,\langle k|\mathcal{O}_{\Pi_3}|l\rangle\,\langle l|c^\dagger_{\sigma'}|i\rangle\ , \nonumber\\
\end{eqnarray}
where the first sum is over the eigenstates and the second over all permutations $\Pi$ of the indices $\{123\}$. We further have defined $\mathcal{O}_1=c_\sigma$, $\mathcal{O}_2=c^\dagger_\sigma$ and $\mathcal{O}_3=c_{\sigma'}$ and e.g. $\Pi_1$ denotes the permutation of the first index. Here the different choice of convention for the Fourier transform simplifies the notation since otherwise the sign of the frequency associated with the creation operator would have to be permuted.
The function $\phi$ is given by the integral
\begin{eqnarray}
\lefteqn{\phi(E_i,E_j,E_k,E_l,\omega_1,\omega_2,\omega_3) =}\nonumber\\
&&\int_0^\beta\! d\tau_1  \int_0^{\tau_1}\! d\tau_2 \int_0^{\tau_2}\! d\tau_3\, e^{-\beta E_i} e^{(E_i-E_j)\tau_1} e^{(E_j-E_k)\tau_2} \times\nonumber\\
&&\times e^{(E_k-E_l)\tau_3} \times e^{i(\omega_1\tau_1+\omega_2\tau_2+\omega_3\tau_3)}\ .
\end{eqnarray}
The latter expression can be evaluated by taking care of the delta functions that arise from equal energies, with the final result
\begin{widetext}
\begin{eqnarray}
&&\phi (E_i,E_j,E_k,E_l,\omega_1,\omega_2,\omega_3)=
\frac{1}{i\omega_3+E_k-E_l} \times\nonumber \\
&&\left[\frac{1-\delta_{\omega_2,-\omega_3}\delta_{E_j,E_l}}{i(\omega_2+\omega_3)+E_j-E_l}\left(
\frac{e^{-\beta E_i}+e^{-\beta E_j}}{i\omega_1+E_i-E_j}-
\frac{e^{-\beta E_i}+e^{-\beta E_l}}{i(\omega_1+\omega_2
+ \omega_3)+E_i-E_l}\right) \right.\nonumber \\
&&\left. +\delta_{\omega_2,-\omega_3}\delta_{E_j,E_l}
\left(\frac{e^{-\beta E_i}+e^{-\beta E_j}}{\left( i\omega_1+E_i-E_j\right)^2}-\beta\frac{e^{-\beta E_j}}{i\omega_1+E_i-E_j}\right) 
-\frac1{i\omega_2+E_j-E_k}  \times \right.\nonumber\\  
&&\left. \left( \frac{e^{-\beta E_i}+e^{-\beta E_j}}{i\omega_1+E_i-E_j}-\left( 1-\delta
_{\omega_1,-\omega_2}\delta_{E_i,E_k}\right)
\frac{e^{-\beta E_i}-e^{-\beta E_k}}{i(\omega_1+\omega
_2)+E_i-E_k}+\beta e^{-\beta E_i}\delta_{\omega
_1,-\omega_2}\delta_{E_i,E_k}\right)\right]\ .
\end{eqnarray}
\end{widetext}
\section{Two-particle vertex in the atomic limit}
\label{app::b}
For the calculations without bath-sites, we have used an explicit expression for the two-particle vertex in the atomic limit. This can be obtained from result of Appendix \ref{app::a} together with the definition of the vertex, Eqn. \ref{eqn::gammanb}. Here we reintroduce explicitly the dependence on $\omega_4$, which yields a more symmetric form of the result. We also switch back to the usual convention for the Fourier transform, for which the energy conservation reads $\omega_1+\omega_3=\omega_2+\omega_4$.

Using that the eigenenergies of $H-\mu\, n$ for the impurity states $|0\rangle$,$|\uparrow\downarrow\rangle$,$|\uparrow\rangle$,$|\downarrow\rangle$ are given by $0,0,U/2,U/2$, respectively, after some algebra the vertex is obtained as
\begin{equation}
\gamma^{\uparrow\uparrow} = \beta\frac{U^2}{4} \frac{\delta_{\omega_1,\omega_2}-\delta_{\omega_2,\omega_3}}{\omega_1^2\omega_3^2}\left(\omega_1^2+\frac{U^2}{4}\right)\left(\omega_3^2+\frac{U^2}{4}\right)\ ,
\end{equation}
\begin{widetext}
\begin{eqnarray}
\gamma^{\uparrow\downarrow} = -U + \frac{U^3}{8} \frac{\omega_1^2+\omega_2^2+\omega_3^2+\omega_4^2}{\omega_1 \omega_2 \omega_3 \omega_4} + \frac{3 U^5}{16 \omega_1 \omega_2 \omega_3 \omega_4} &+& \beta\frac{U^2}{4} \frac{1}{1+e^{\beta U/2}} \frac{2\delta_{\omega_2,-\omega_3}+\delta_{\omega_1,\omega_2}}{\omega_2^2 \omega_3^2}\left(\omega_2^2+\frac{U^2}{4}\right)\left(\omega_3^2+\frac{U^2}{4}\right)\nonumber\\
&-& \beta\frac{U^2}{4} \frac{1}{1+e^{-\beta U/2}} \frac{2\delta_{\omega_2,\omega_3}+\delta_{\omega_1,\omega_2}}{\omega_1^2 \omega_3^2}\left(\omega_1^2+\frac{U^2}{4}\right)\left(\omega_3^2+\frac{U^2}{4}\right)\ .\nonumber\\
\end{eqnarray}
\end{widetext}

\bibliography{spert.bib}

\begin{thebibliography}{17}
\expandafter\ifx\csname natexlab\endcsname\relax\def\natexlab#1{#1}\fi
\expandafter\ifx\csname bibnamefont\endcsname\relax
  \def\bibnamefont#1{#1}\fi
\expandafter\ifx\csname bibfnamefont\endcsname\relax
  \def\bibfnamefont#1{#1}\fi
\expandafter\ifx\csname citenamefont\endcsname\relax
  \def\citenamefont#1{#1}\fi
\expandafter\ifx\csname url\endcsname\relax
  \def\url#1{\texttt{#1}}\fi
\expandafter\ifx\csname urlprefix\endcsname\relax\def\urlprefix{URL }\fi
\providecommand{\bibinfo}[2]{#2}
\providecommand{\eprint}[2][]{\url{#2}}

\bibitem[{\citenamefont{Hewson}(1993)}]{hewson}
\bibinfo{author}{\bibfnamefont{A.~C.} \bibnamefont{Hewson}},
  \emph{\bibinfo{title}{The Kondo Problem to Heavy Fermions}}
  (\bibinfo{publisher}{Cambridge Univ. Press}, \bibinfo{address}{Cambridge},
  \bibinfo{year}{1993}).

\bibitem[{\citenamefont{Savkin et~al.}(2005)\citenamefont{Savkin, Rubtsov,
  Katsnelson, and Lichtenstein}}]{savkin}
\bibinfo{author}{\bibfnamefont{V.~V.} \bibnamefont{Savkin}},
  \bibinfo{author}{\bibfnamefont{A.~N.} \bibnamefont{Rubtsov}},
  \bibinfo{author}{\bibfnamefont{M.~I.} \bibnamefont{Katsnelson}},
  \bibnamefont{and} \bibinfo{author}{\bibfnamefont{A.~I.}
  \bibnamefont{Lichtenstein}}, \bibinfo{journal}{Phys. Rev. Lett.}
  \textbf{\bibinfo{volume}{94}}, \bibinfo{eid}{026402}
  (pages~\bibinfo{numpages}{4}) (\bibinfo{year}{2005}).

\bibitem[{\citenamefont{Kotliar et~al.}(2006)\citenamefont{Kotliar, Savrasov,
  Haule, Oudovenko, Parcollet, and Marianetti}}]{kotliar}
\bibinfo{author}{\bibfnamefont{G.}~\bibnamefont{Kotliar}},
  \bibinfo{author}{\bibfnamefont{S.~Y.} \bibnamefont{Savrasov}},
  \bibinfo{author}{\bibfnamefont{K.}~\bibnamefont{Haule}},
  \bibinfo{author}{\bibfnamefont{V.~S.} \bibnamefont{Oudovenko}},
  \bibinfo{author}{\bibfnamefont{O.}~\bibnamefont{Parcollet}},
  \bibnamefont{and} \bibinfo{author}{\bibfnamefont{C.~A.}
  \bibnamefont{Marianetti}}, \bibinfo{journal}{Rev. Mod. Phys.}
  \textbf{\bibinfo{volume}{78}}, \bibinfo{eid}{865}
  (pages~\bibinfo{numpages}{87}) (\bibinfo{year}{2006}).

\bibitem[{\citenamefont{Katsnelson et~al.}(2008)\citenamefont{Katsnelson,
  Irkhin, Chioncel, Lichtenstein, and de~Groot}}]{katsnelson:rmp}
\bibinfo{author}{\bibfnamefont{M.~I.} \bibnamefont{Katsnelson}},
  \bibinfo{author}{\bibfnamefont{V.~Y.} \bibnamefont{Irkhin}},
  \bibinfo{author}{\bibfnamefont{L.}~\bibnamefont{Chioncel}},
  \bibinfo{author}{\bibfnamefont{A.~I.} \bibnamefont{Lichtenstein}},
  \bibnamefont{and} \bibinfo{author}{\bibfnamefont{R.~A.}
  \bibnamefont{de~Groot}}, \bibinfo{journal}{Rev. Mod. Phys.}
  \textbf{\bibinfo{volume}{80}}, \bibinfo{eid}{315}
  (pages~\bibinfo{numpages}{64}) (\bibinfo{year}{2008}).

\bibitem[{\citenamefont{Dai et~al.}(2005)\citenamefont{Dai, Haule, and
  Kotliar}}]{dai:045111}
\bibinfo{author}{\bibfnamefont{X.}~\bibnamefont{Dai}},
  \bibinfo{author}{\bibfnamefont{K.}~\bibnamefont{Haule}}, \bibnamefont{and}
  \bibinfo{author}{\bibfnamefont{G.}~\bibnamefont{Kotliar}},
  \bibinfo{journal}{Phys. Rev. B} \textbf{\bibinfo{volume}{72}},
  \bibinfo{eid}{045111} (pages~\bibinfo{numpages}{6}) (\bibinfo{year}{2005}).

\bibitem[{\citenamefont{Rubtsov et~al.}(2005)\citenamefont{Rubtsov, Savkin, and
  Lichtenstein}}]{rubtsovctqmc}
\bibinfo{author}{\bibfnamefont{A.~N.} \bibnamefont{Rubtsov}},
  \bibinfo{author}{\bibfnamefont{V.~V.} \bibnamefont{Savkin}},
  \bibnamefont{and} \bibinfo{author}{\bibfnamefont{A.~I.}
  \bibnamefont{Lichtenstein}}, \bibinfo{journal}{Phys. Rev. B}
  \textbf{\bibinfo{volume}{72}}, \bibinfo{eid}{035122}
  (pages~\bibinfo{numpages}{9}) (\bibinfo{year}{2005}).

\bibitem[{\citenamefont{Werner et~al.}(2006)\citenamefont{Werner, Comanac, de'
  Medici, Troyer, and Millis}}]{werner1}
\bibinfo{author}{\bibfnamefont{P.}~\bibnamefont{Werner}},
  \bibinfo{author}{\bibfnamefont{A.}~\bibnamefont{Comanac}},
  \bibinfo{author}{\bibfnamefont{L.}~\bibnamefont{de' Medici}},
  \bibinfo{author}{\bibfnamefont{M.}~\bibnamefont{Troyer}}, \bibnamefont{and}
  \bibinfo{author}{\bibfnamefont{A.~J.} \bibnamefont{Millis}},
  \bibinfo{journal}{Phys. Rev. Lett.} \textbf{\bibinfo{volume}{97}},
  \bibinfo{eid}{076405} (pages~\bibinfo{numpages}{4}) (\bibinfo{year}{2006}).

\bibitem[{\citenamefont{Haule}(2007)}]{haule}
\bibinfo{author}{\bibfnamefont{K.}~\bibnamefont{Haule}},
  \bibinfo{journal}{Phys. Rev. B} \textbf{\bibinfo{volume}{75}},
  \bibinfo{eid}{155113} (pages~\bibinfo{numpages}{12}) (\bibinfo{year}{2007}).

\bibitem[{\citenamefont{Rubtsov et~al.}(2008)\citenamefont{Rubtsov, Katsnelson,
  and Lichtenstein}}]{rubtsov}
\bibinfo{author}{\bibfnamefont{A.~N.} \bibnamefont{Rubtsov}},
  \bibinfo{author}{\bibfnamefont{M.~I.} \bibnamefont{Katsnelson}},
  \bibnamefont{and} \bibinfo{author}{\bibfnamefont{A.~I.}
  \bibnamefont{Lichtenstein}}, \bibinfo{journal}{Phys. Rev. B}
  \textbf{\bibinfo{volume}{77}}, \bibinfo{eid}{033101}
  (pages~\bibinfo{numpages}{4}) (\bibinfo{year}{2008}).

\bibitem[{\citenamefont{Hafermann et~al.}(2007)\citenamefont{Hafermann, Brener,
  Rubtsov, Katsnelson, and Lichtenstein}}]{hafermann}
\bibinfo{author}{\bibfnamefont{H.}~\bibnamefont{Hafermann}},
  \bibinfo{author}{\bibfnamefont{S.}~\bibnamefont{Brener}},
  \bibinfo{author}{\bibfnamefont{A.~N.} \bibnamefont{Rubtsov}},
  \bibinfo{author}{\bibfnamefont{M.~I.} \bibnamefont{Katsnelson}},
  \bibnamefont{and} \bibinfo{author}{\bibfnamefont{A.~I.}
  \bibnamefont{Lichtenstein}}, \bibinfo{journal}{JETP Lett.}
  \textbf{\bibinfo{volume}{86}}, \bibinfo{pages}{677} (\bibinfo{year}{2007}).

\bibitem[{\citenamefont{Brener et~al.}(2008)\citenamefont{Brener, Hafermann,
  Rubtsov, Katsnelson, and Lichtenstein}}]{brener}
\bibinfo{author}{\bibfnamefont{S.}~\bibnamefont{Brener}},
  \bibinfo{author}{\bibfnamefont{H.}~\bibnamefont{Hafermann}},
  \bibinfo{author}{\bibfnamefont{A.~N.} \bibnamefont{Rubtsov}},
  \bibinfo{author}{\bibfnamefont{M.~I.} \bibnamefont{Katsnelson}},
  \bibnamefont{and} \bibinfo{author}{\bibfnamefont{A.~I.}
  \bibnamefont{Lichtenstein}}, \bibinfo{journal}{Phys. Rev. B}
  \textbf{\bibinfo{volume}{77}}, \bibinfo{eid}{195105}
  (pages~\bibinfo{numpages}{12}) (\bibinfo{year}{2008}).

\bibitem[{\citenamefont{Toschi et~al.}(2007)\citenamefont{Toschi, Katanin, and
  Held}}]{toschi}
\bibinfo{author}{\bibfnamefont{A.}~\bibnamefont{Toschi}},
  \bibinfo{author}{\bibfnamefont{A.~A.} \bibnamefont{Katanin}},
  \bibnamefont{and} \bibinfo{author}{\bibfnamefont{K.}~\bibnamefont{Held}},
  \bibinfo{journal}{Phys. Rev. B} \textbf{\bibinfo{volume}{75}},
  \bibinfo{eid}{045118} (pages~\bibinfo{numpages}{8}) (\bibinfo{year}{2007}).

\bibitem[{\citenamefont{Georges et~al.}(1996)\citenamefont{Georges, Kotliar,
  Krauth, and Rozenberg}}]{georges}
\bibinfo{author}{\bibfnamefont{A.}~\bibnamefont{Georges}},
  \bibinfo{author}{\bibfnamefont{G.}~\bibnamefont{Kotliar}},
  \bibinfo{author}{\bibfnamefont{W.}~\bibnamefont{Krauth}}, \bibnamefont{and}
  \bibinfo{author}{\bibfnamefont{M.}~\bibnamefont{Rozenberg}},
  \bibinfo{journal}{Rev. Mod. Phys.} \textbf{\bibinfo{volume}{68}},
  \bibinfo{pages}{13} (\bibinfo{year}{1996}).

\bibitem[{\citenamefont{Koch et~al.}(2008)\citenamefont{Koch, Sangiovanni, and
  Gunnarsson}}]{koch}
\bibinfo{author}{\bibfnamefont{E.}~\bibnamefont{Koch}},
  \bibinfo{author}{\bibfnamefont{G.}~\bibnamefont{Sangiovanni}},
  \bibnamefont{and}
  \bibinfo{author}{\bibfnamefont{O.}~\bibnamefont{Gunnarsson}}
  (\bibinfo{year}{2008}), \bibinfo{note}{arXiv:0804.3320}.

\bibitem[{\citenamefont{Pairault et~al.}(2000)\citenamefont{Pairault,
  S\'en\'echal, and Tremblay}}]{tremblay}
\bibinfo{author}{\bibfnamefont{S.}~\bibnamefont{Pairault}},
  \bibinfo{author}{\bibfnamefont{D.}~\bibnamefont{S\'en\'echal}},
  \bibnamefont{and} \bibinfo{author}{\bibfnamefont{A.-M.~S.}
  \bibnamefont{Tremblay}}, \bibinfo{journal}{Eur. Phys. J. B}
  \textbf{\bibinfo{volume}{16}}, \bibinfo{pages}{85} (\bibinfo{year}{2000}).

\bibitem[{\citenamefont{Baym and Kadanoff}(1961)}]{BK}
\bibinfo{author}{\bibfnamefont{G.}~\bibnamefont{Baym}} \bibnamefont{and}
  \bibinfo{author}{\bibfnamefont{L.~P.} \bibnamefont{Kadanoff}},
  \bibinfo{journal}{Phys. Rev.} \textbf{\bibinfo{volume}{124}},
  \bibinfo{pages}{287} (\bibinfo{year}{1961}).

\bibitem[{\citenamefont{Jarrell and Gubernatis}(1996)}]{jarrell}
\bibinfo{author}{\bibfnamefont{M.}~\bibnamefont{Jarrell}} \bibnamefont{and}
  \bibinfo{author}{\bibfnamefont{J.~E.} \bibnamefont{Gubernatis}},
  \bibinfo{journal}{Phys. Rep.} \textbf{\bibinfo{volume}{269}},
  \bibinfo{pages}{133} (\bibinfo{year}{1996}).

\end{thebibliography}

\end{document}